\documentclass[conference]{IEEEtran}
\IEEEoverridecommandlockouts

\usepackage{cite}
\usepackage{amsmath,amssymb,amsfonts}
\usepackage{algorithmic}
\usepackage{graphicx}
\usepackage{textcomp}
\usepackage{xcolor}
\usepackage{multirow}
\usepackage{url}
\usepackage{hyperref}

\usepackage{lipsum}
\usepackage{fancyhdr}
\fancypagestyle{sec}{\lhead{\tmpx}}

\fancypagestyle{arXiv}
{
   \fancyhf{}
   \chead{\textcolor{gray}{This paper has been accepted to the 35th International Conference on Field-Programmable Logic \\ and Applications (FPL 2025), 1-5 September 2025, Leiden, The Netherlands.}}
   \cfoot{ \fontsize{=9pt}{10pt}\selectfont  \textcolor{gray}{© 2025 IEEE. Personal use of this material is permitted. Permission from IEEE must be obtained for all other uses, in any current or future media, including reprinting/republishing this material for advertising or promotional purposes, creating new collective works, for resale or redistribution to servers or lists, or reuse of any copyrighted component of this work in other works}\\\fontsize{=10pt}{12pt}\selectfont\thepage}
   
}

\def\BibTeX{{\rm B\kern-.05em{\sc i\kern-.025em b}\kern-.08em
    T\kern-.1667em\lower.7ex\hbox{E}\kern-.125emX}}
\begin{document}

\title{SparseDPD: A Sparse Neural Network-based Digital Predistortion FPGA Accelerator for RF Power Amplifier Linearization\\
}

\author{\IEEEauthorblockN{Manno Versluis, Yizhuo Wu, Chang Gao}

\IEEEauthorblockA{Department of Microelectronics, Delft University of Technology, The Netherlands}}

\maketitle

\begin{abstract}
Digital predistortion (DPD) is crucial for linearizing radio frequency (RF) power amplifiers (PAs), improving signal integrity and efficiency in wireless systems. Neural network (NN)-based DPD methods surpass traditional polynomial models but face computational challenges limiting their practical deployment. This paper introduces SparseDPD, an FPGA accelerator employing a spatially sparse phase-normalized time-delay neural network (PNTDNN), optimized through unstructured pruning to reduce computational load without accuracy loss. Implemented on a Xilinx Zynq-7Z010 FPGA, SparseDPD operates at 170 MHz, achieving exceptional linearization performance (ACPR: -59.4 dBc, EVM: -54.0 dBc, NMSE: -48.2 dB) with only 241 mW dynamic power, using 64 parameters with 74\% sparsity. This work demonstrates FPGA-based acceleration, making NN-based DPD practical and efficient for real-time wireless communication applications. Code is publicly available at \url{https://github.com/MannoVersluis/SparseDPD}.
\end{abstract}


\section{Introduction}
\thispagestyle{arXiv}
Wireless communication systems form the foundation of modern connectivity, supporting applications ranging from mobile networks to satellite links. Central to these systems are radio frequency (RF) power amplifiers (PAs), which amplify signals to sufficient power levels for reliable transmission over long distances. However, RF PAs exhibit nonlinear behavior, especially when driven near their saturation region to optimize power efficiency~\cite{Wesemann}. This nonlinearity introduces unwanted distortion, degrading signal integrity and reducing system efficiency.

Digital predistortion (DPD) has become a widely adopted technique to counteract these distortions. DPD preprocesses the input signal with an inverse model of the PA’s nonlinearity, effectively linearizing the output. Traditional DPD approaches, such as those using memory polynomials~\cite{1703853}, have been favored for their simplicity and reasonable performance. Yet, as wireless standards evolve, requiring higher data rates, wider bandwidths, and complex modulation schemes, these methods face limitations.

NN-based DPDs~\cite{Wang2019,Chen2023TWC,10176354} have recently emerged as a superior alternative, leveraging the ability of NNs to model complex nonlinear PA behaviors. However, they incur significant computational overhead, posing challenges for real-time low-power deployment to meet industry needs. FPGAs, with their reprogrammability and parallel processing strengths, are ideal for accelerating DPD. However, previous FPGA-based DPD implementations mostly focused on polynomial models, leaving NN-based approaches underexplored.

This paper presents SparseDPD, an FPGA accelerator tailored for sparse NN-based DPD. 
Our design utilizes a spatially sparse PNTDNN to deliver high throughput, low power consumption, low utilization, and exceptional linearization performance. 
By applying unstructured pruning~\cite{Han2015NIPS}, we reduce the model size by 74\% without sacrificing accuracy. 
Implemented on a Xilinx Zynq-7Z010 FPGA, SparseDPD achieves a peak clock frequency of 170\,MHz. 
When evaluated with a 20-MHz-bandwidth 64-QAM signal, it attains an ACPR of -59.4\,dBc, an EVM of -54.0\,dBc, and an NMSE of -48.2\,dB, all with a dynamic power consumption of just 241\,mW. 
This work demonstrates that FPGA acceleration can bridge the gap between the computational demands of NN-based DPD and the real-time needs of wireless systems, offering a practical and efficient solution for next-generation communication networks.

\begin{figure}[t]
    \centering
    \includegraphics[width=0.5\textwidth]{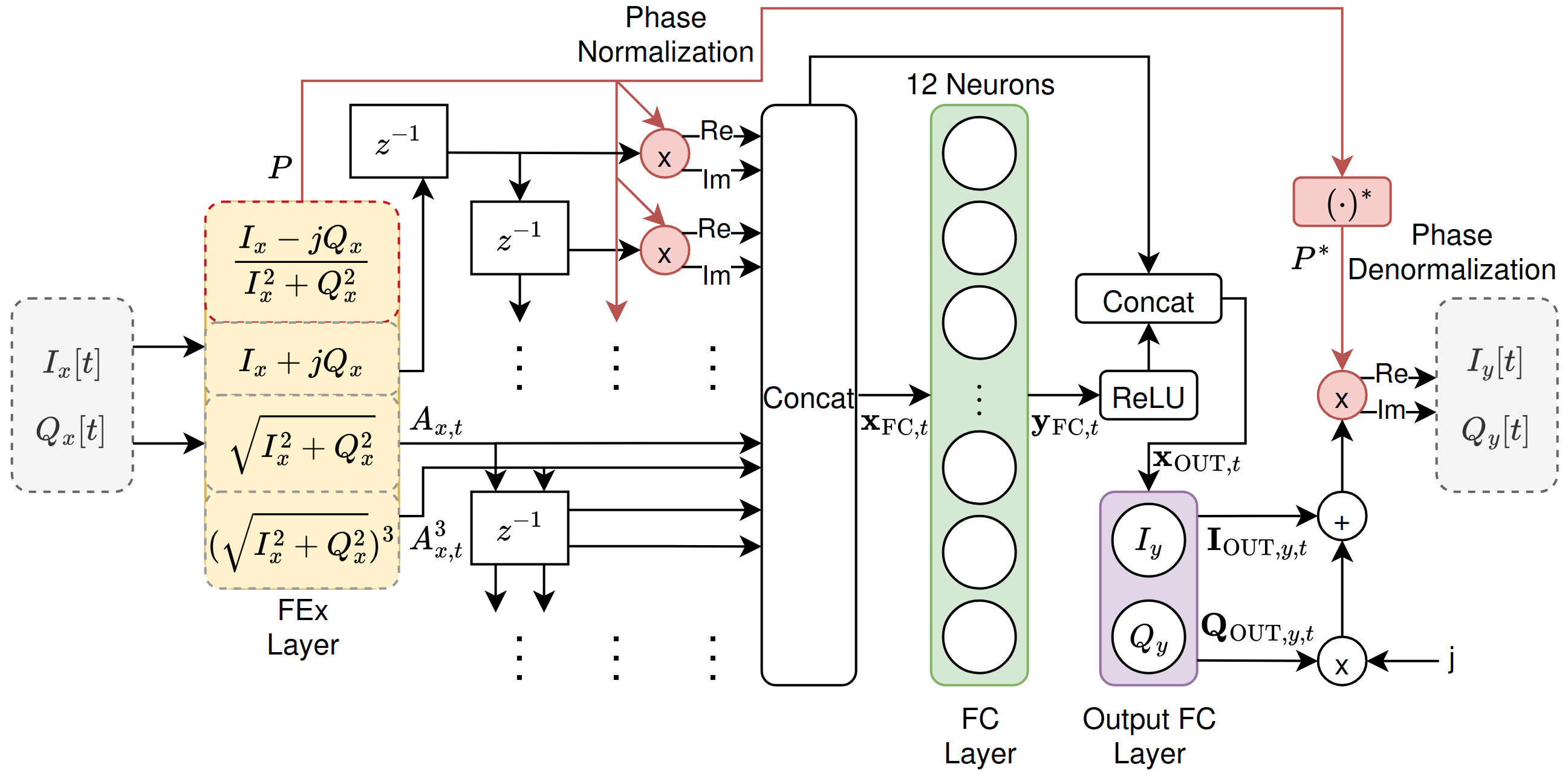}
    \caption{The PNTDNN architecture, showing the flow from input signals ($I_x$, $Q_x$) through feature extraction, phase normalization, fully connected layers, and phase denormalization to produce the predistorted output ($I_y$, $Q_y$).}
    \label{fig:PNTDNN_architecture}
\end{figure}

\section{Algorithm} 
The PNTDNN model, illustrated in Figure~\ref{fig:PNTDNN_architecture} and detailed in \cite{10176354}, comprises four stages: feature extraction, phase normalization, fully connected (FC) layers, and phase denormalization, that preprocess the input signal to mitigate PA nonlinearities.
\subsection{Feature Extraction Layer}
The feature extraction (FEx) stage processes the in-phase ($I_{x,t}$) and quadrature ($Q_{x,t}$) components of the input signal at time $t$, constructing the feature vector $\mathbf{x}_{\text{FEx},t}$ as follows:
\begin{align}
    \mathbf{x}_{\text{FEx},t} &= \begin{bmatrix}
    \frac{I_{x,t} - j Q_{x,t}}{A_{x,t}}, I_{x,t} + j Q_{x,t}, A_{x,t}, A_{x,t}^3
    \end{bmatrix} 
    \label{FEx_eq} \\
    A_{x,t} &= \sqrt{I_{x,t}^2 + Q_{x,t}^2}
    \label{eq:A}
\end{align}
\noindent where $A_{x,t}$ is the signal amplitude. This vector captures both linear and nonlinear signal characteristics, with $A_{x,t}^3$ specifically addressing higher-order distortions critical for PA compensation.

\subsection{Time Delay}
To incorporate the PA’s memory effects, this stage introduces delayed versions of the feature extraction outputs, spanning a memory depth of $n$ timesteps. The following vectors are defined:
\begin{align}
    \mathbf{K}_{x,t} &= [I_{x,t-1}+jQ_{x,t-1}, ..., I_{x,t-n}+jQ_{x,t-n}] \\
    \mathbf{A}_{x,t} &= [A_{x,t}, ..., A_{x,t-n}] \\
    \mathbf{A}_{x,t}^3 &= [A^3_{x,t}, ..., A^3_{x,t-n}]
\end{align}
These time-delayed vectors provide temporal context, enabling the model to address time-dependent nonlinearities in subsequent layers.

\subsection{Phase Normalization}
The phase normalization stage aligns the phases of the delayed inputs $\mathbf{K}_{x,t}$ by subtracting the phase of the current input, facilitating processing in the FC layers. The normalized output is computed as:
\begin{align}\mathbf{I}_{\text{PN\_out},x,t}+j\mathbf{Q}_{\text{PN\_out},x,t} &= (\mathbf{K}_{x,t})\times P \\
    P &= \frac{I_{x,t}-jQ_{x,t}}{\sqrt{I_{x,t}^2+Q_{x,t}^2}}
    \label{eq:P}
\end{align}
\noindent where $P$ is a complex scalar that rotates the delayed inputs. The current timestep is excluded from normalization, as it would redundantly yield $A_{x,t} + 0j$.

\subsection{FC Layers}
The phase-normalized outputs ($\mathbf{I}_{\text{PN\_out},x,t}$, $\mathbf{Q}_{\text{PN\_out},x,t}$) and amplitude vectors ($\mathbf{A}_{x,t}$, $\mathbf{A}_{x,t}^3$) are concatenated to form the input to the FC layers:
\begin{align}
    \mathbf{x}_{\text{FC},t} &= [\mathbf{I}_{\text{PN\_out},x,t}, \mathbf{Q}_{\text{PN\_out},x,t}, \mathbf{A}_{x,t}, \mathbf{A}^3_{x,t}]
\end{align}
The FC layers then process this input through a sequence of transformations:
\begin{align}
    \mathbf{y}_{\text{FC},t} &= \mathbf{W}_{\text{FC}}\mathbf{x}_{\text{FC},t} + \mathbf{b}_{\text{FC}} \\
    \mathbf{x}_{\text{OUT},t} &= [\mathbf{x}_{\text{FC},t}, \text{ReLU}(\mathbf{y}_{\text{FC},t})] \\
    \mathbf{y}_{\text{OUT},t} &= \mathbf{W}_{\text{OUT}}\mathbf{x}_{\text{OUT},t} + \mathbf{b}_{\text{OUT}} \\
    [I_{\text{OUT},y,t}, Q_{\text{OUT},y,t}] &= \mathbf{y}_{\text{OUT},t}
\end{align}
\noindent where $\mathbf{W}_{\text{FC}}$, $\mathbf{b}_{\text{FC}}$, $\mathbf{W}_{\text{OUT}}$, and $\mathbf{b}_{\text{OUT}}$ are the trainable weights and biases of the first and output FC layers, respectively. The ReLU activation introduces nonlinearity, enabling the network to model complex PA distortions.
\begin{figure}[!t]
    \centering
    \includegraphics[width=0.35\textwidth]{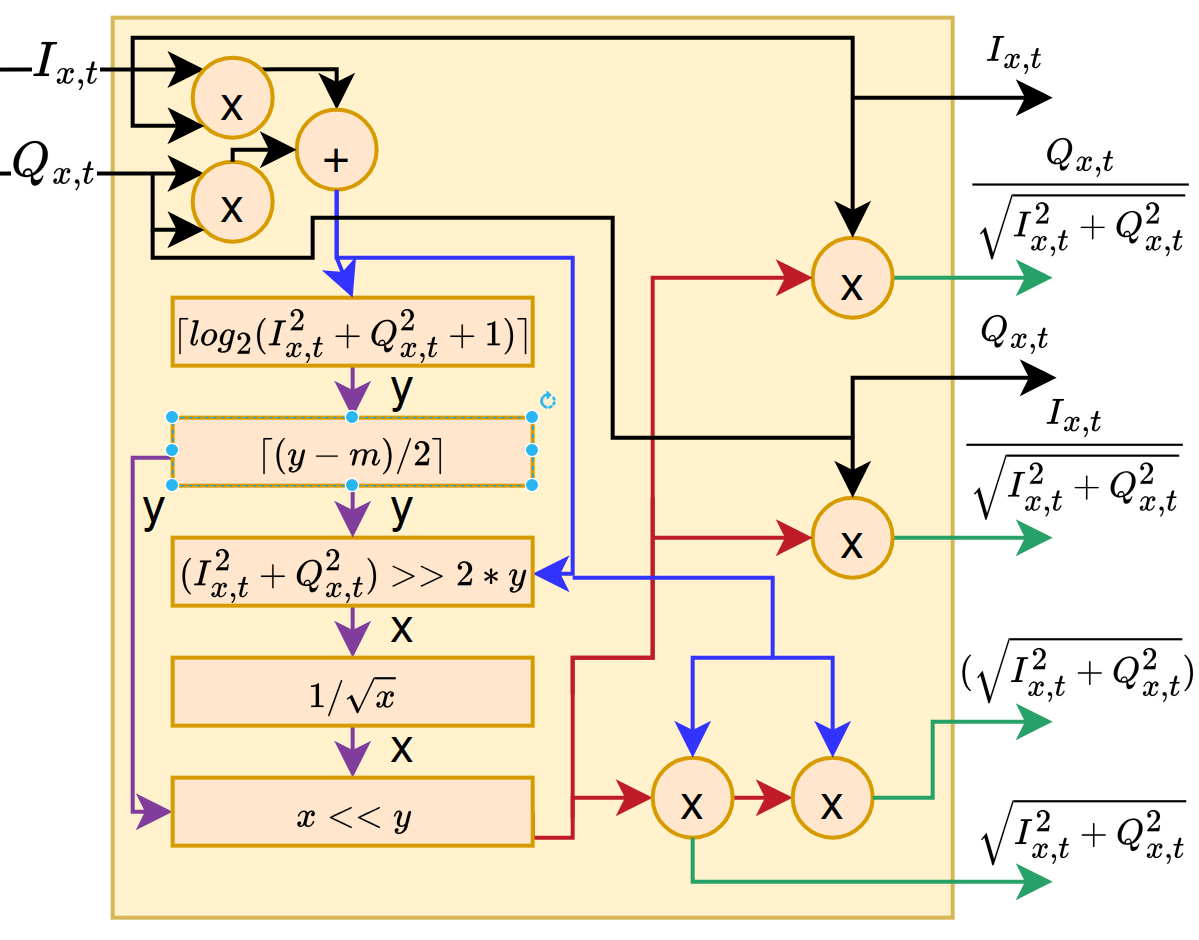}
    \caption{The FEx layer implementation, where m is the window size of the $1/\sqrt{x}$ input.}
    \label{fig:FEx diagram}
\end{figure}
\begin{figure}[!t]
    \centering
    \includegraphics[width=0.25\textwidth]{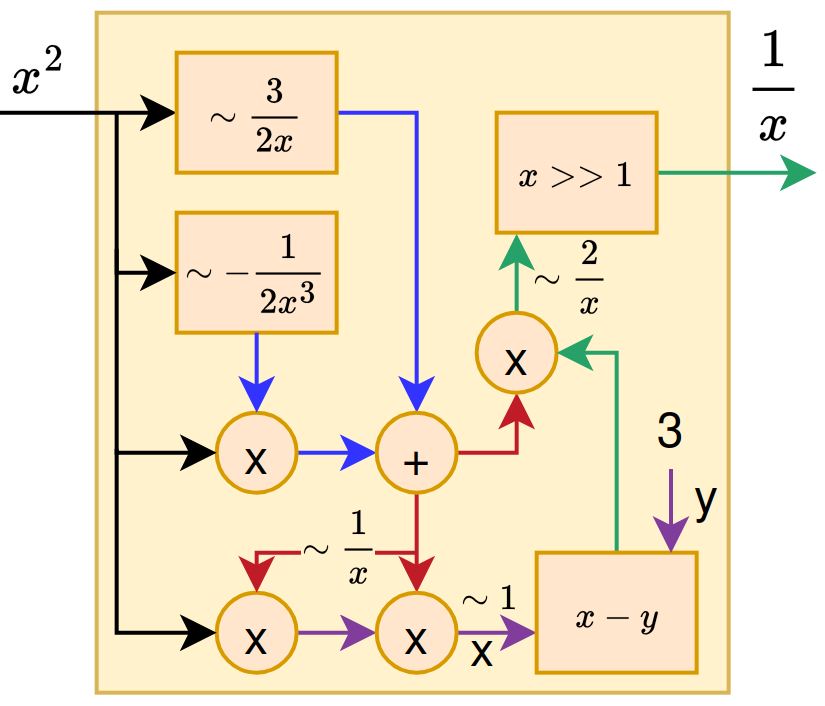}
    \caption{The inverse square root implementation.}
    \label{fig:inv sqrt diagram}
\end{figure}
\begin{figure*}
    \centering
    \includegraphics[width=1.0\textwidth]{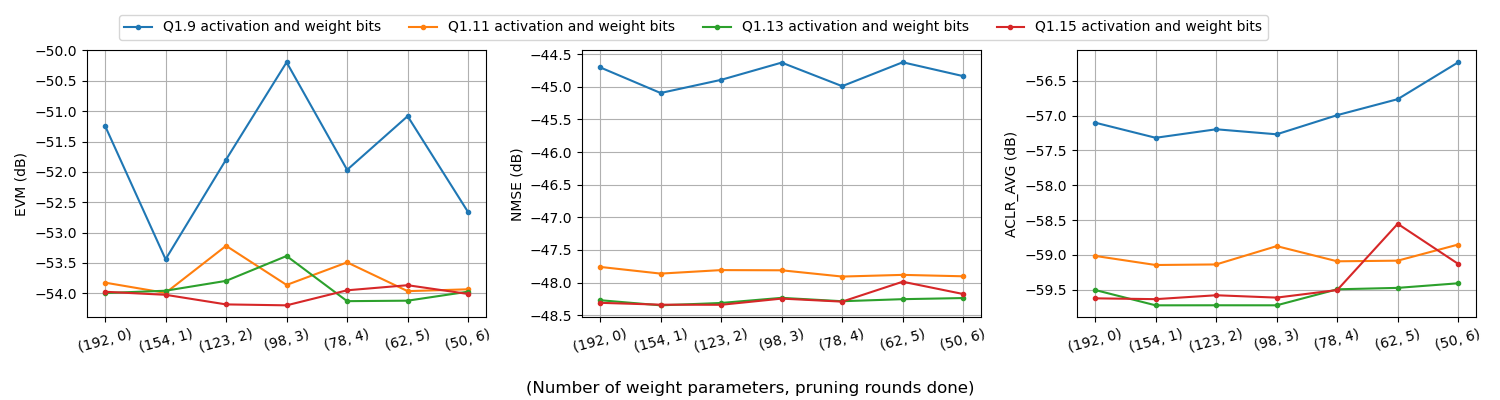}
    \caption{DPD model accuracy while pruning for various amounts of bits, with 12 bias parameters and a hidden size of 12, one seed only. Trained for 400 epochs before pruning and 200 epochs after each pruning iteration.}
    \label{fig:PNTDNN pruning rounds plot}
\end{figure*}
\subsection{Phase Denormalization}
The final stage restores the original phase to the FC layer outputs:
\begin{equation}
    I_{y,t}+jQ_{y,t} = (I_{\text{OUT},y,t}+jQ_{\text{OUT},y,t})P^*
    \label{eq:phase denorm}
\end{equation}
\noindent where $P^*$ is the complex conjugate of $P$ from \eqref{eq:P}, ensuring the predistorted signal aligns with the input’s phase properties.

\section{Accelerator Design}
The SparseDPD accelerator is engineered for efficiency on the FPGA, optimizing each PNTDNN component for fixed-point arithmetic and parallel processing to maximize throughput and minimize latency.

\subsection{Feature Extraction}
The feature extraction layer, illustrated in Figure~\ref{fig:FEx diagram}, $I_x$ and $Q_x$ input components to compute the inverse square root $1/\sqrt{I_x^2 + Q_x^2}$, a computationally demanding operation. To address this, we implement an approximation method using a lookup table (LUT) to provide initial estimates ($1.5x^{(0)}$ and $0.5(x^{(0)})^3$), followed by two Newton-Raphson iterations \cite{Parhami_2000}:
\begin{equation}
    x^{(n+1)} = 0.5x^{(n)}(3-z(x^{(n)})^2)
    \label{inv sqrt approx iter}
\end{equation} 
where $z = I_x^2 + Q_x^2$ and $x$ approximates $1/\sqrt{z}$. This approach significantly reduces computational complexity while preserving sufficient precision for DPD tasks, as shown in Figure \ref{fig:inv sqrt diagram}.

\subsubsection{Square Root Approximation}
A critical optimization exploits a property of square roots: for an $N$-bit input, the least significant $N/2$ bits can alter the output by at most 1. This stems from the relationship:
\begin{equation}
    x+1 = \sqrt{x^2+2x+1}
    \label{sqrt_N}
\end{equation}
\noindent where $x^2 = N$, implying that increasing the square root by 1 requires an input increment of $2\sqrt{N} + 1$. Leveraging this, we right-shift the input by $2a$ bits until the most significant nonzero bit lies within a predefined window, compute the inverse square root using this reduced input, and left-shift the result by $a$ bits. This technique halves the input size to the inverse square root module, introducing a minor inaccuracy that neural network training compensates for, unlike truncation, which would give an input of 0 for small inputs. This optimization is essential because the Zynq-7Z010 FPGA’s DSP blocks support 25×18-bit multiplications, while $I_x^2 + Q_x^2$ for 14-bit inputs produces 28 bits, exceeding the DSP capacity.

\subsection{Fully Connected Layers and Pruning}
The FC layers are implemented by multiplying neuron inputs with their weights in parallel. The results, combined with biases, are aggregated using a carry-save adder (CSA) tree, followed by an addition to generate each neuron’s output.

To optimize resource usage, we apply unstructured pruning, setting insignificant weights to zero, eliminating their multiplications, and reducing the CSA tree’s input width accordingly.

\subsection{Fixed-point Data Representation}
To ensure hardware efficiency while maintaining model accuracy, we adopt a 14-bit Q1.13 fixed-point format (1 integer bit, 13 fractional bits, two’s complement) for NN activations, weights, and input I/Q data. FC layer outputs are clamped to the [-1, 1) range, with intermediate CSA tree computations quantized to Q2.13 to detect overflow. The I/Q output data uses a 29-bit Q2.27 fixed-point format to preserve precision after phase denormalization, as rounding to 14 bits degrades performance. Quantization Aware Training, implemented using MP-DPD~\cite{10502240}, minimizes accuracy loss due to fixed-point quantization, ensuring robust linearization performance.

\begin{table*}
    \caption{Comparison table of DPD Hardware Accelerator Implementations}
    \centering
    \resizebox{\textwidth}{!}{
    \begin{tabular}{c|c c c c c c c c|c c c c}
    \hline
    & \multicolumn{8}{c}{\multirow{2}{*}{\textbf{Platform Characteristics and Performance}}} & \multicolumn{4}{|c}{\textbf{Signal Bandwidth and}}\\
    & \multicolumn{8}{c}{} & \multicolumn{4}{|c}{\textbf{Quality of DPD+PA outputs}}\\
    \hline
    & \multirow{2}{*}{Architecture} & \multirow{2}{*}{Model} & Weight/Act. & \multirow{2}{*}{\#Param} & \multirow{2}{*}{$\frac{Operations}{Sample}$} & \multirow{2}{*}{$f_{clk}$} & \multirow{2}{*}{$f_{s,I/Q}$} & Total & \multirow{2}{*}{$f_{BB}$} & \multirow{2}{*}{ACPR} & \multirow{2}{*}{EVM} & \multirow{2}{*}{NMSE}\\
    & & & Precision & & & & & Power & & & &\\
    &  &  & (bits) &  &  & (MHz) & (MSps) & (W) & (MHz) & (dBc) & (dBc) & (dB)\\
    \hline
    This work & FPGA (7Z010) & PNTDNN & W14A14 & 64 & 72 & 170 & 170 & 0.405 & 20 & -59.4 & -54.0 & -48.2\\
    \cite{Sparse-NN-DPD} & FPGA (ZCU111) & ETDNN & W20A16 & 40 & - & - & 368.64 & - & 46.08 & $\sim$-50 & $\sim$-42.3& $\sim$-39.5\\
    \cite{9762118} & FPGA (XC7Z020) & MP & W?A16 & 9 & 30 & 250 & 250 & 0.244 & 20 & -49 & - & -\\
    \cite{9653793} & FPGA (XC7VX485T) & GMP & W?A16 & 38 & 149 & - & 400 & 0.89 & 100 & -46.5 & - & -38.5\\
    \cite{9099463} & FPGA (ZCU102) & GMP & W?A16 & 36 & 17 & 300 & 2400 & 0.96 & 400 & -44.7 & -39.2 & -\\
    \cite{GPU_pruning} & GPU & TDNN & FP32 & 909 & $\sim$1,818 & $\sim$2,300 & 1000 & $\leq$320 & 200 & -45.2 & -35.3 & -38.3\\
    \cite{DPD-NeuralEngine} & ASIC & RNN & W12A12 & 502 & 1,026 & 2,000 & 250 & 0.2 & 60 & -45.3 & -39.8 & -\\
    \hline
    \end{tabular}
    }
    \label{comparison table}
\end{table*}

\section{Results and Discussion}
\subsection{Experimental setup}
\begin{figure}
    \centering
    \includegraphics[width=0.75\linewidth]{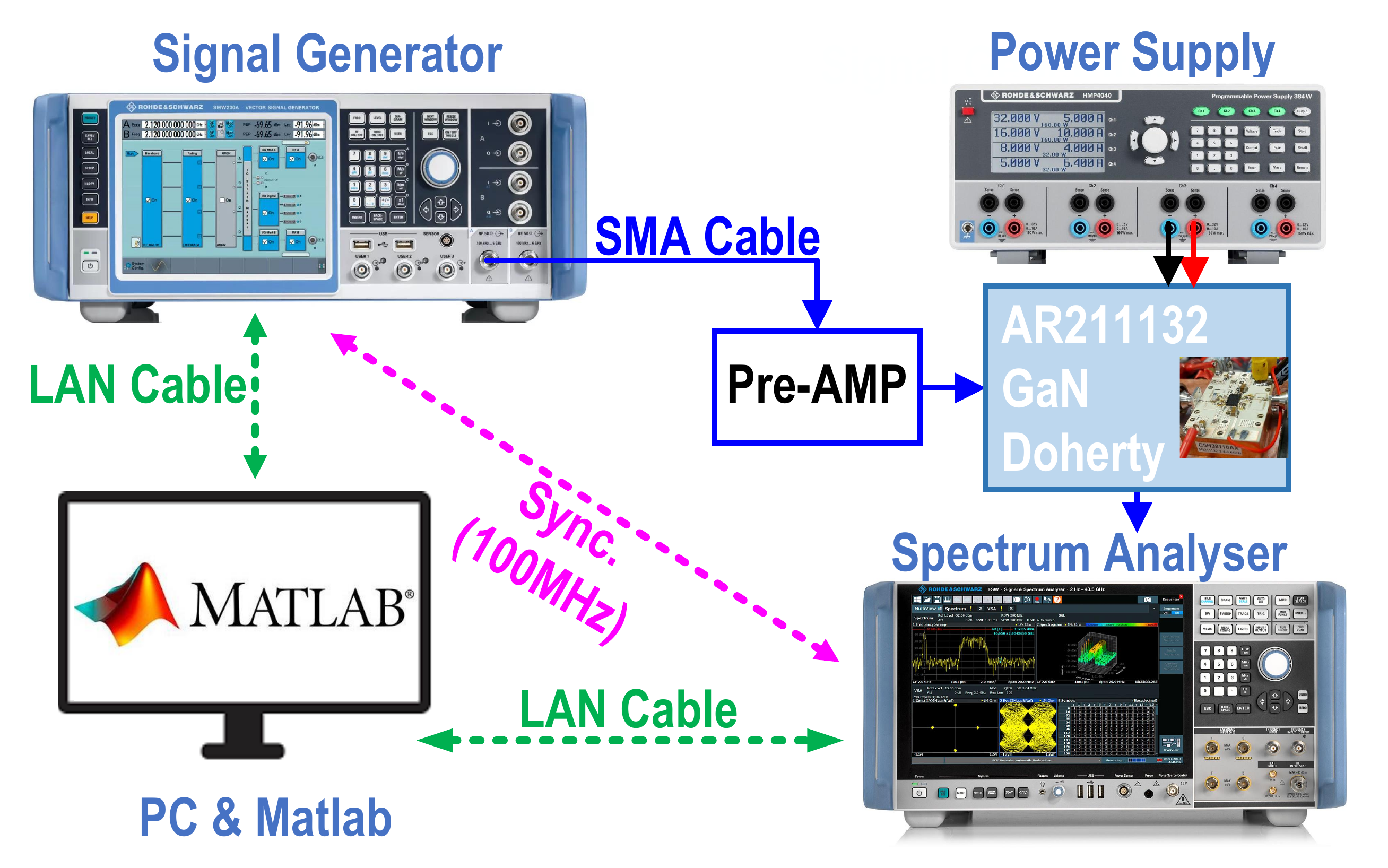}
    \caption{Experimental setup for dataset acquisition and DPD performance measurement. }
    \label{fig:platform}
\end{figure}

The PNTDNN model, with memory depth $n = 2$ and hidden size 12, was evaluated using the \texttt{OpenDPD} framework~\cite{Wu_2024} with simulated results. The dataset comprised 172,035 samples of a 20 MHz 64-QAM signal, including MATLAB-generated ideal baseband inputs and corresponding PA outputs from a 3.5 GHz GaN Doherty PA operating at 41.5 dBm without DPD. As shown in Fig.~\ref{fig:platform}, the setup involved generating and up-converting the baseband signal in MATLAB, transmitting via R\&S SMW200A, passing through the PA, then down-converting and digitizing using R\&S FSW43. Data was split 60/20/20 for training/validation/testing. QAT was conducted for 400 epochs, with subsequent pruning iterations, each comprising 200 epochs of QAT. Training used \texttt{ReduceLROnPlateau} with initial learning rate $10^{-3}$ (reset after pruning iterations), minimum $10^{-6}$, batch size 256, frame length 500, stride 1, patience 6, decay factor 0.5, and 6 pruning rounds removing the lowest 20\% magnitude weights each. The FPGA design was simulated on Xilinx Zynq-7Z010, with power consumption (Table~\ref{power report}) analyzed using switching-activity-annotated post-implementation simulations.

\subsection{Model Accuracy and Throughput}
Fig.~\ref{fig:PNTDNN pruning rounds plot} compares model accuracy across different bit precisions and pruning iterations. 
The results demonstrate that increasing activation and weight precision beyond 14 bits provides no additional accuracy gains, while reducing precision below 12 bits greatly decreases performance. 
This analysis, based on one seed, confirms the 14-bit fixed-point representation as an optimal balance between accuracy and efficiency.

The SparseDPD accelerator achieves a throughput of 1 sample per clock cycle at 170\,MHz, equivalent to 12.2\,GOPS, with further pipelining difficult. 
Throughput could be further enhanced by using multiple model instances on the FPGA, sharing FEx layer outputs to process multiple samples concurrently. 
However, this was not implemented due to the size constraints of the Zynq-7Z010 FPGA.
\subsection{FPGA Implementation}
Resource utilization for the FPGA implementation is detailed in Table~\ref{FPGA resource utilization table} and power utilization in Table~\ref{FPGA power utilization table}. The FEx layer leverages block RAM (BRAM) and LUTs to approximate the terms \( 1.5x^{(0)} \) and \( 0.5(x^{(0)})^3 \), where \( x = 1/\sqrt{I_x^2 + Q_x^2} \), in the inverse square root computation. Less accurate approximations could reduce resource demands but would either compromise model accuracy or require an additional Newton-Raphson iteration (Eq.~\ref{inv sqrt approx iter}), increasing DSP usage and latency. With DSP utilization already at 82.5\%, this trade-off was avoided to maintain performance and efficiency.

\begin{table}
    \centering
    \caption{Power utilization on the Zynq-7Z010 FPGA, }
    \resizebox{0.3\textwidth}{!}{\begin{tabular}{c c|c}
    Static & Total & 0.164 W\\
    \hline
    & Clocks & 0.023 W\\
    & Signals & 0.065 W\\
    Dynamic & Logic & 0.037 W\\
    & BRAM & 0.040 W\\
    & DSP & 0.077 W\\
    \hline
    Dynamic & Total & 0.241 W\\
    \hline
    Static + Dynamic& Total & 0.405 W\\
    \end{tabular}}
    \label{power report}
    \label{FPGA power utilization table}
\end{table}

\begin{table}
    \centering
    \caption{Resource utilization on Zynq-7Z010 FPGA, }
    \resizebox{0.45\textwidth}{!}{\begin{tabular}{c|c c c c c}
    \hline
    & LUT & FF & DSP & BRAM\\
    \hline
    Available & 17600 & 35200 & 80 & 60\\
    Total used & 2298 & 1724 & 66 & 13\\
    & (13.1\%) & (4.9\%) & (82.5\%) & (21.7\%)\\
    FEx layer & 536 & 561 & 10 & 13\\
    Phase Norm & 2 & 0 & 2 & 0\\
    Phase Denorm & 0 & 0 & 4 & 0\\
    FC layer & 874 & 680 & 26 & 0\\
    ReLU & 70 & 0 & 0 & 0\\
    Output FC layer & 798 & 104 & 21 & 0\\
    Shift register & 0 & 70 & 0 & 0\\
    \hline
    \end{tabular}}
    \label{FPGA resource utilization table}
\end{table}
\subsection{Comparison with Previous Work}
Table~\ref{comparison table} benchmarks the SparseDPD accelerator against other DPD hardware implementations. 
The table contains designs with varying bandwidths due to limited DPD hardware accelerator implementation publications. 
Our design achieves superior linearization performance, with an ACPR of -59.4\,dBc, an EVM of -54.0\,dBc, and an NMSE of -48.2\,dB at a 20\,MHz baseband frequency. 
While some prior works report higher sampling rates, SparseDPD’s architecture supports parallelization to increase throughput, limited only by FPGA resource availability. Compared to traditional memory polynomial models, which benefit from fewer parameters due to their fixed structures, SparseDPD maintains a compact 64-parameter footprint with 74\% weight sparsity from unstructured pruning, far fewer than other NN-based implementations, except for~\cite{Sparse-NN-DPD}, which has a similar amount of parameters using structured pruning, though with lower linearization performance compared to our method.

\section{Conclusion}
We presented SparseDPD, an FPGA accelerator for neural network-based digital predistortion, which uses a pruned phase-normalized time-delay neural network to deliver -59.4\,dBc ACPR, -54.0\,dBc EVM, and -48.2\,dB NMSE. Running on an Xilinx Zynq-7Z010 FPGA at 170 MHz with 241 mW dynamic power, promising to support efficient, real-time DPD for next-generation wireless systems. 
Future work would be to use a higher BW dataset with a larger model, or to exploit by adopting more types of sparsity~\cite{Liu2022,Wu2025} in the neural network model for further computational and memory access saving.

\section*{Acknowledgment}
This work was partially supported by the European Research Executive Agency (REA) under the Marie Skłodowska-Curie Actions (MSCA) Postdoctoral Fellowship program, Grant No. 101107534 (AIRHAR).

\bibliographystyle{IEEEtran}
\bibliography{IEEEabrv,paper_bibliography}
\end{document}